\documentclass[10pt,conference]{IEEEtran}

\usepackage[dvipsnames,usenames]{xcolor}
\usepackage{cite}
\usepackage{makecell}

\usepackage[normalem]{ulem}
\usepackage{syntax}
\usepackage[frozencache]{minted}
\setminted{fontsize=\footnotesize,linenos,breaklines=true}
\usepackage{makecell}
\usepackage{balance}
\usepackage{graphicx}
\usepackage{graphics}
\usepackage{stfloats}
\usepackage{tabularx}
\usepackage{hyperref}
\usepackage{booktabs}
\usepackage{multirow}

\usepackage[breakable, theorems, skins]{tcolorbox}

\usepackage{tabu}

\DeclareRobustCommand{\mybox}[2][gray!15]{%
\begin{tcolorbox}[   %
        breakable,
        left=0pt,
        right=0pt,
        top=0pt,
        bottom=0pt,
        colback=#1,
        colframe=#1,
        width=\columnwidth, 
        enlarge left by=0mm,
        boxsep=5pt,
        arc=0pt,outer arc=0pt,
        ]
        #2
\end{tcolorbox}
}

\begin{document}
\title{SQLRepair: Identifying and Repairing Mistakes in Student-Authored SQL Queries}

\author{\IEEEauthorblockN{Kai Presler-Marshall, Sarah Heckman, Kathryn T. Stolee}
\IEEEauthorblockA{North Carolina State University \\
Raleigh, North Carolina \\
Email: \{kpresle, sarah\_heckman, ktstolee\}@ncsu.edu}
}

\maketitle

\thispagestyle{plain}
\pagestyle{plain}

\begin{abstract}

Computer science educators seek to understand the types of mistakes that students make when learning a new (programming) language so that they can help students avoid those mistakes in the future.  While educators know what mistakes students regularly make in languages such as C and Python, students struggle with SQL and regularly make mistakes when working with it.  We present an analysis of mistakes that students made when first working with SQL, classify the types of errors introduced,  and provide suggestions on how to avoid them going forward.  In addition, we present an automated tool, SQLRepair, that is capable of repairing errors introduced by undergraduate programmers when writing SQL queries.  
Our results show that students find repairs produced by our tool comparable in understandability to queries written by themselves or by other students, suggesting that SQL repair tools may be useful in an educational context.  We also provide to the community a benchmark of SQL queries written by  the students in our study that we used for evaluation of SQLRepair.

\end{abstract}

\IEEEpeerreviewmaketitle

\section{Introduction}

Understanding how beginners work with a new programming language and the types of mistakes that they make can help instructors better tailor their lesson plans to avoid previous pitfalls~\cite{uml, testingmistakes}.  We consider SQL, a widely-used language for interacting with relational databases.  SQL is taught in many undergraduate computer science programs~\cite{sqlteaching, sqllearning}, but may not be part of the core curriculum.  It is regularly used by professional and amateur developers alike~\cite{sosurvey}, including those with little formal computer science background~\cite{datasciencecurriculum,datascienceundergrad}. 

While the types of mistakes that students make when working with languages such as C and Java are relatively well studied~\cite{novicemistakes,studentmistakes,javamistakes}, we know less about mistakes made in special-purpose languages such as SQL.  We seek to understand the types of mistakes that undergraduate students, who are relatively familiar with %
Java, make when working with SQL.  Understanding these mistakes can help educators ensure that they have the resources necessary to support computer science students and end-user programmers alike, which may include automated support~\cite{apreducation}.

In addition to an analysis of student mistakes, we propose a tool, SQLRepair, which can automatically fix some of the errors students introduce.\footnote{We adopt terminology used in existing work on SQL education: students make a \textit{mistake} while solving a problem, introducing one or more \textit{error}s into the query.  Note that this diverges from terminology frequently used in testing literature where the term would be \textit{fault} instead of \textit{error}.  We choose \textit{error} for consistency with existing work.}  While there are tools for automated repair of programs in languages such as C and Java~\cite{simfix,learningcode,angelix,weimer2009}, to the best of our knowledge, no existing techniques attempt to repair errors in SQL queries.  Our repair process first attempts non-synthesis repair based on a predefined ruleset.  As needed, it uses a satisfiability modulo theory (SMT) solver~\cite{z3} to further synthesize repairs.  

We frame our work around the following research questions:
\begin{itemize}
    \item \textbf{RQ1}: What types of mistakes do beginners make when working with SQL?
    \item \textbf{RQ2}: How well can SQLRepair fix errors introduced by beginning SQL programmers?
    \item \textbf{RQ3:} Do students find SQLRepair-repaired queries to be more understandable than queries written by other students?
\end{itemize}

To answer our research questions, we conducted an empirical evaluation to understand student mistakes (RQ1), evaluate SQLRepair's ability to repair the errors in the student-written queries (RQ2), and determine the repair quality (RQ3).
Students in two undergraduate computer science courses at a large public university in the United States, North Carolina State University (NCSU), were given a short introduction to SQL and then asked to write queries to solve problems associated with a sample database.  
For each problem, students were provided an example \texttt{(source, destination)} table pair that demonstrated the desired transformation (similar to programming by example (PBE) techniques)~\cite{gulwanipbe} and were asked to write a SQL query that would complete the transformation. Incorrect queries were followed by additional examples (up to three) to demonstrate the intended behavior.  Any SQL query that did not correctly solve the problem was analyzed for errors and considered a candidate for repair.  Students were then asked to evaluate up to four human-written or tool-generated queries, judging each for understandability.  
Our work makes the following contributions:
\begin{itemize}
    \item quantitative and qualitative classifications of the types of errors introduced by beginning SQL programmers, 
    \item a tool capable of repairing 29.1\% of the observed errors in SQL queries, %
    \item a benchmark dataset of realistic SQL errors  gathered from undergraduate computer science students, and
    \item a demonstration that tool-repaired SQL queries are equal in understandability to human-written queries.
\end{itemize}

\section{Study}
\label{sec:dataset}
To provide a dataset for analyzing mistakes (RQ1) and evaluating SQLRepair (RQ2, RQ3), we conducted a two-phase study with students from two undergraduate computer science courses.  %
In Phase~1,  we conducted a study with students from the Summer 2019 offering of a 2\textsuperscript{nd}-year Java programming course.  This phase demonstrated the viability of our approach, gave us preliminary data for RQ1 and RQ2, and motivated additional enhancements to our tool.  In Phase~2, we put repairs produced by SQLRepair directly in front of students to understand whether our tool-generated repairs are understandable (RQ3).   Students were recruited from the Fall 2020 offerings of a 2\textsuperscript{nd}-year Java programming course and a 3\textsuperscript{rd}-year Software Engineering course.  Our tool and instructions on how to set it up are available.\footnote{\url{http://github.com/kpresler/sqlrepair}}

\subsection{Phase~1}
\label{sec:phase1}

We collected a dataset of SQL queries written by introductory programmers to understand the type of mistakes students make by analysing the errors they introduce, and ascertain SQLRepair's ability to repair the errors.

\subsubsection{Design}
\label{sec:design}

Eighteen students were given a lecture on SQL functionality and syntax, including compound select queries, various datatypes, \mintinline{sql}{JOIN}, \mintinline{sql}{COUNT}, \mintinline{sql}{DISTINCT}, and \mintinline{sql}{GROUP BY}.  Students were informed that we were interested in studying how beginners work with SQL and the types of mistakes that they make.  Next, they were given a ten-problems to solve; each problem had a \texttt{(source, destination)} table pair and students were asked to write a SQL query that would accomplish the transformation.  Each problem had two or three pairs of \texttt{(source, destination)} tables that acted as test cases that must be passed simultaneously for the query to be considered correct.  The major concept of each problem is shown in Table~\ref{tab:problems}.  For example, the major concept introduced in Problem 10 was grouping, and there were two sets of \texttt{(source, destination)} table pairs for evaluating the query.  The problems and data used were based on the UMLS dataset, a health and biomedical vocabulary dataset made available free-of-charge by the NIH, which was chosen for offering a large amount of structured data~\cite{umlsdataset}. %

\begin{table}[]
\caption{
    Major concept in each problem and the total number of \texttt{(source, destination)} tables in the problem specifications.
}
\begin{tabular}{clc}
Problem & Major Concept                       & Number of Table Pairs       \\
\toprule
1       & Single-condition select                       & 3          \\
2       & Select with projection       & 2 \\
3       & Inequality              & 3          \\
4       & Projection and inequality        & 2 \\
5       & Compound select  & 2 \\
6       & Compound select with AND & 2      \\
7       & Distinct                            & 2 \\
8       & Ordering                            & 2 \\
9       & Joins                       & 2 \\
10      & Grouping                            & 2      
\end{tabular}

\label{tab:problems}
\end{table}

Students were shown one \texttt{(source, destination)} table pair at a time.  Each student received a paper handout that contained the first pair for each problem. To avoid learning effects, the problems were given in a random order. Students submitted their queries into a web application. If the application detected that the first  pair had been solved successfully, the query was then tested against subsequent pairs. If a query failed a subsequent pair, that pair was revealed to the student. Students spent approximately 40 minutes working on all problems and were reminded every ten minutes to move on to the next problem if they had been stuck for more than five minutes.  Students were compensated with participation credit.%

\begin{figure}
    \includegraphics[width=1\linewidth]{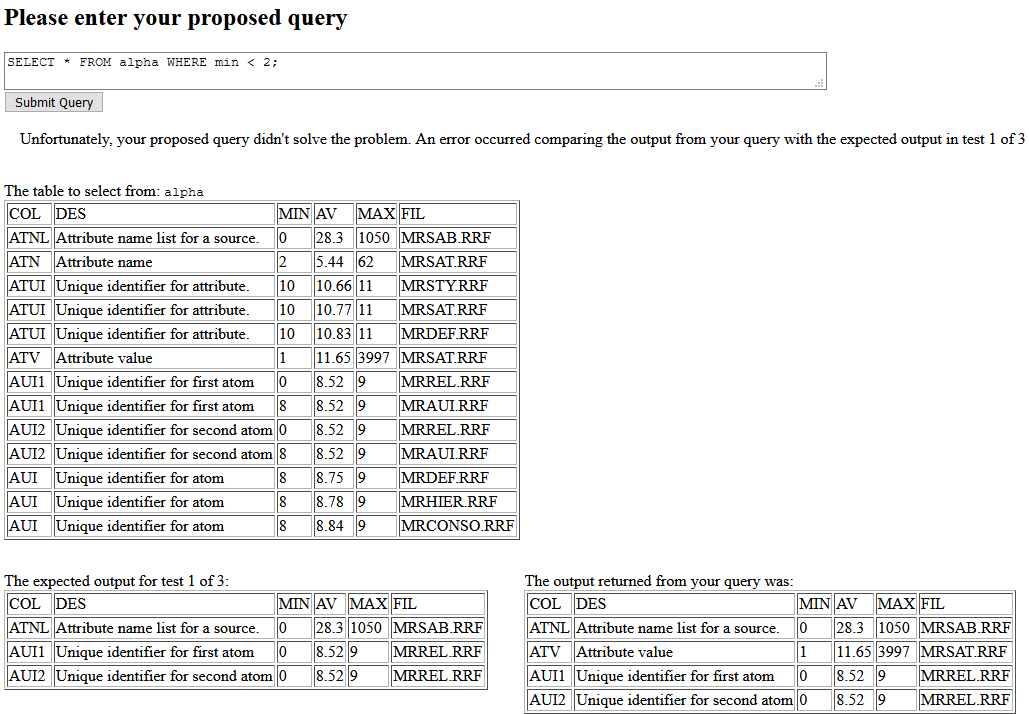}
    \vspace{-6pt}
    \caption{
        The application for students to submit  SQL queries.
    }
    \label{fig:sqlsearcher}
    \vspace{-12pt}
\end{figure}

The web application  is shown in Figure~\ref{fig:sqlsearcher}.  In this example, a student submitted the query \mintinline{sql}{SELECT * FROM alpha WHERE min < 2;}, which was incorrect, as communicated through the message, \emph{``Unfortunately, your proposed query didn't solve the problem \dots"}; the actual output from executing the query is shown alongside the expected output (destination table).  If the query produces the correct output for all table pairs, the student was congratulated and told to move on to the next problem.  The application records the participant's unique ID, submission time, proposed query, and whether the problem was solved correctly or not.  At the end of the study, students completed a brief demographics survey, which asked questions such as their prior programming experience, their experience with SQL, and whether they had any comments on the introduction to SQL or the problems themselves.

\subsubsection{Participants}
We recruited participants from a 2\textsuperscript{nd}-year Java programming course (CS2).  CS2 is the second computer science course taken by majors and minors at NCSU. By this point, students are exposed to programming in Java.  Eighteen students from the Summer 2019 offering of CS2 participated, but only 12 students submitted one or more SQL queries as part of the study.  Of the 12 active participants, three identified as female.  Ten students said they had three or fewer years of programming experience (min: 0, max: 8, average: 2.6) and none had more than a year of professional programming experience.  One student reported prior database experience.

\subsubsection{Dataset}
We collected 362 SQL queries written by 12 different students.  Of these, 35 were correct.  Of the 327 incorrect queries, 124 had syntax error(s) and 203 had semantic error(s).  Students submitted between 7 and 65 queries (average: 32.2, median: 28.5).  Students made between one and 21 attempts per problem (average: 4.6, median: 3.5) and attempted between two and ten problems (average and median:~6.5).

\subsection{Phase~2}
\label{sec:phase2}

In Phase~2, we build on Phase~1 and further evaluate SQLRepair by putting repaired queries directly in front of students to assess query quality.%

\subsubsection{Design}

Phase~2 was similar to Phase~1 in that students were given the same introductory SQL lecture and the same set of problems to solve.  However, some changes were made to the study format and content, as follows:

Due to the COVID-19 pandemic, Phase~2 was performed online via Zoom.  After the introduction to SQL and the study, each participant was assigned to an individual breakout room to work in for the remainder of the session.  To ensure that each participant was engaged and working, the first author rotated between each room at least once to answer any technical questions that arose.  Students could also use Zoom's ``Ask for help'' functionality to request assistance. 

While the study problems were identical to Phase~1, we made operational changes to suit the online format:
\begin{itemize}
    \item Instead of a paper handout, each student received the randomly ordered problems as a PDF.
    \item Instead of students entering their participant ID manually, the web application automatically included each student's random ID in each problem submission.
    \item The post-study demographics survey was converted from a paper handout to a Google Form.  Students were asked to include their participant ID in their submission.
    
\end{itemize}

\noindent Additionally, after composing queries for a problem, students evaluated the understandability of several solution queries for that problem (Section~\ref{sec:evaluatingsqlrepair}).

\subsubsection{Evaluating SQLRepair} \label{sec:evaluatingsqlrepair}
We wanted students to assess the understandability of tool-repaired queries by comparing them against human-written queries. As a majority of software engineering effort is spent on maintenance~\cite{seFallacies}, we consider understandability, as a proxy for ease of maintenance, to be paramount.  We seek a minimally-invasive way of gathering information on students' program comprehension as they evaluate queries without the feeling of being watched~\cite{pairProgramming}.  Thus, we opt for short surveys deployed after each question and separately at the end of the study.

First, we populated a database with data from Phase~1, giving us 29 unique correct queries and 19 unique repaired queries SQLRepair produced from incorrect queries.  
Next, we modified the web application to use SQLRepair to attempt to repair incorrect queries that students wrote during the study.  We did this through brief post-problem surveys: after solving each problem, students were asked to rate the understandability of up to four different queries using a modified Likert scale, with 1 indicating the query was very difficult to understand and 7 that it was very easy to understand.  As an alternate workflow, after making at least five attempts at a problem over at least five minutes, %
students were presented with an ``I'm tired of this problem'' button.  Upon clicking it, they would be given the voting options shown, despite having never solved the problem correctly.

The four possible queries presented to students were:
\begin{itemize}
    \item \textbf{MyCorrectQuery}: A correct query written by the student (available if they solved the problem correctly).
    \item \textbf{MyRepairedQuery}: A repair of an incorrect query written by the student (available if they got the problem wrong at least once, and SQLRepair was able to repair one of their queries.\footnote{Incorrect queries were considered starting with the most recent incorrect submission, and repairs were attempted until a query was successfully repairable, or, to ensure sufficient responsiveness of the web application, the repair process had failed ten times.})
    \item \textbf{OtherCorrectQuery}: A correct query written by someone else (a participant from Phase~1 of the study; a query from this category was always available).
    \item \textbf{OtherRepairedQuery}: A repair of an incorrect query written by someone else (a participant from Phase~1 of the study; a query from this category was always available).
    
\end{itemize}

\begin{figure}
    \includegraphics[width=1\linewidth]{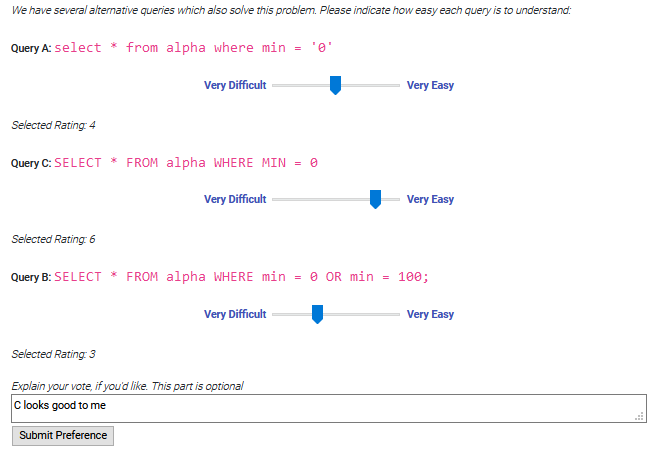}
    \vspace{-6pt}
    \caption{
        An example of how students voted on the understandability of queries.
    }
    \label{fig:votingscreen}
    \vspace{-12pt}
\end{figure}

The queries were labeled A through D, and presented in a random order.  An example with three queries is shown in Figure~\ref{fig:votingscreen}.  For queries written by others, query selection was pseudo-random: each query was associated with a count of how many times it had been shown to a student for voting, and each time a query was needed for voting, the application selected the query with the smallest vote count. Identical queries were consolidated (for instance, if the first and fourth queries were identical, the query would only appear once).

\subsubsection{Participants}

In Fall 2020, we distributed recruitment emails to students in two undergraduate courses: CS2 and a 3\textsuperscript{rd}-year Software Engineering course (SE).  SE is a fifth-semester course, and by this point, students have been exposed to Java, C, x86 assembly, and JavaScript.  Additionally, prior to our study, the SE students received an in-class lecture on SQL, although not hands-on practice with it.  Students in both classes were invited to sign up for one of four two-hour virtual lab sessions held.  In all, 104 students signed up to participate in a session; 71 students attended and participated for at least thirty minutes.  The first of four sessions was used as a pilot  for the improved SQLRepair tool and new format. Feedback was collected and  data from this group was discarded.  Participants from Phase~1 could not participate in Phase~2.

Seventy-three students from CS2 signed up; 46 ultimately participated.  Thirty-one students from SE signed up; 24 ultimately participated.  After discarding data from the pilot study, we retained data from 33 CS2 and 19 SE students.  Students in CS2 reported up to seven years of prior programming experience (average and median: 2); students in SE reported up to 8 years (average: 5, median: 4).  Sixteen participants from CS2 and five from SE identified as female.

\subsubsection{Dataset}

We collected 2,420 SQL queries from 52 students.  Of these, 216 were correct; of the 2,204 incorrect queries, 693 had syntax error(s) and 1511 had semantic error(s).  Students submitted between 1 and 118 queries (average: 42.4, median: 37.5).  Students attempted between 1 and 10 problems (average and median: 7) and made between 1 and 50 attempts per problem (average: 6.4, median:~4).

The 33 students from CS2 submitted 1,660 queries.  Of these, 122 were correct; of the 1,538 incorrect queries, 556 had syntax error(s) and 982 had semantic error(s).  Students submitted a median of 41 queries (max: 118) and attempted between 1 and 10 problems (average: 6.8, median: 7.5).%

The 19 students from SE submitted 760 queries.  Of these, 94 were correct; of the 666 incorrect queries, 137 had syntax error(s) and 529 had semantic error(s).  Students submitted a median of 36 queries (max:  79).  Students attempted between 1 and 10 problems (average: 7.4, median: 7).%

\begin{table}[tb]
 \centering
\caption{
    A breakdown of all of the queries submitted.
}
\begin{tabular}{lrrr}
 & \multicolumn{2}{c}{\textbf{Course}}  &                                                                           \\
 & CS2             & SE & \makecell{\textbf{Total}}   \\
\midrule
Correct        & 157                    & 94  &  251           \\
Syntax Error   & 680                    & 137 &  817           \\
Semantic Error & 1,185                  & 529 &  1,714           \\
      
\bottomrule
\makecell{\textbf{Total}} & 2,022 & 760 & 2,782  \\
\end{tabular}
\label{tab:breakdown}
\end{table}

\subsection{Data Summary}  A summary of all of the queries collected across both phases of our study, and their correctness or error category, is shown in Table~\ref{tab:breakdown}.  We performed a Mann-Whitney test between the two CS2 courses (Summer 2019 and Fall 2020 from Phase~1 and Phase~2, respectively) looking for significant differences on successes per problem.  Our analysis revealed  that the differences between them were not statistically significant ($p=.31$), so the data from both were combined for further analysis.  The data from SE remained significantly different ($p=.0016$) and was kept separate.

\subsection{Analysis}
\label{sec:sqleval}

We use the errors that students introduce into SQL queries they write as a proxy for the mistakes made while solving the problem.  To identify student mistakes for RQ1, we executed each student-written query against the  source and destination tables using a MySQL 5.7 database.  
Any query where the database returned an error message was considered to have syntax error(s).\footnote{This understates the number of SQL syntax errors as  MySQL 5.7 supports functionality not part of the official SQL specification, such as wrapping strings in double quotes or using operators such as \mintinline[breaklines]{sql}{&&} instead of \mintinline[breaklines]{sql}{AND}.}  For the remaining queries, we compared the actual output table to the expected output for the problem. When they were different, the query was considered to have semantic error(s).

To identify syntax error categories, we manually grouped queries with similar errors together.  For example, students submitted the queries: 

\begin{minted}[breaklines,linenos=false]{sql}
SELECT CUI1, RUI FROM bravo where REL='RO', 'SY';
SELECT CUI1, RUI, FROM bravo WHERE CUI2 == C0364349;
\end{minted}

\noindent Both queries have an extra comma, so were grouped together.  We continued this process for all queries with a syntax error.  If there were three or more queries in a category, we gave the category a name. Categories with fewer than three were grouped together into a miscellaneous category.  %

For semantic errors, we manually investigated the query and the output table it produced and grouped together queries with similar errors. For example, students submitted the queries:
\begin{minted}[linenos=false]{sql}
SELECT LAT FROM juliett;
SELECT LAT, STT FROM juliett
\end{minted}  
\noindent Both queries return only a subset of the columns expected (\mintinline[breaklines]{sql}{LAT, STT, ISPREF}) so they were grouped together.  A miscellaneous category was created by grouping together all categories with less than three queries.  %

A single query can contain multiple errors (for instance, a broken operator and unquoted string literal) so some queries were counted for multiple categories.  However, when classifying errors, a single query could be counted towards either the syntax error category \textit{or} semantic error category, but not both.

\section{SQLRepair}
\label{sec:implementation}
SQLRepair follows the correct-by-construction approach to automated program repair~\cite{aprsurvey}.  %
The subset of supported SQL includes queries with compound \mintinline[breaklines]{sql}{WHERE} clauses, integer and string datatypes, \mintinline[breaklines]{sql}{ORDER BY}, and \mintinline[breaklines]{sql}{DISTINCT}.%

\begin{figure}[tb]
\begin{footnotesize}
    \begin{tabular}{lllll}
        item    & price & quantity      & country & seller \\
        \toprule
        apples  & 7     & 500           & US       & Joe's Fruits     \\
        bananas & 3     & 400           & MX       & Nancy's Produce  \\
        oranges & 11    & 300           & MA       & Ahmed's Fruits   \\
        grapes  & 1     & 200           & US       & Raj's Vinyard 
    \end{tabular}

\vspace{12pt}

        \begin{tabular}{llll}
        item   & price & quantity & country \\
        \toprule
        grapes & 1     & 200      & US      
    \end{tabular}
\end{footnotesize}
    \caption{
         Example source (top) and destination (bottom) tables.
         \label{tab:ex-src}    
     }
\end{figure}

To explain how SQLRepair builds constraints from the \texttt{(source, destination)} tables and SQL query, consider the following example.  A user of SQLRepair submits the source  and destination tables shown in Figure~\ref{tab:ex-src} and the SQL query \mintinline[breaklines]{sql}{SELECT * FROM fruitSellers WHERE country=US && quantity < 800}. %
SQLRepair proceeds in two steps: (1) non-synthesis repair, and (2) synthesis repair.

\subsection{Non-Synthesis Repair}
SQLRepair attempts three types of non-synthesis repair over the following types of errors: operator mismatches that result in parse errors, column mismatches that can cause an otherwise correct query to be incorrect, and string repair where a string literal shows up without proper quotes. 

\subsubsection{Operator Mismatch}
\label{sec:opmismatch}
SQLRepair replaces any C/Java-style operators in the provided query with their SQL equivalent.  For example, C/Java use \mintinline{java}{==} for equality checks and \mintinline{java}{&&} for logical AND.  SQL uses \mintinline{sql}{=} and \mintinline{sql}{AND}, respectively.  SQLRepair thus replaces operators such as these.  In the example,  \mintinline[breaklines]{sql}{&&} is replaced with  \mintinline[breaklines]{sql}{AND}, giving us the query, \mintinline[breaklines]{sql}{SELECT * FROM fruitSellers WHERE country=US AND quantity < 800}.

\subsubsection{Column Mismatch} 
\label{sec:colmismatch}
SQLRepair attempts to repair any issues with the column list prior to the \mintinline[breaklines]{sql}{WHERE} clause.  When a column does not exist, a syntax error occurs.  However, column mismatch does not always start with a syntax error.  In the the running example, the source table has five columns while the destination table only has four; however, the SQL query has a \mintinline[breaklines]{sql}{SELECT *} clause, SQLRepair detects and fixes this mismatch.  Thus, the query is updated to \mintinline[breaklines]{sql}{SELECT item, price, quantity, country FROM fruitSellers WHERE country=US AND quantity < 800}. In addition to correcting the column list following \mintinline{sql}{SELECT}, SQLRepair can also rename columns to match the destination table using \mintinline{sql}{AS}. 

\subsubsection{String Repair}
\label{sec:strrepair}
SQLRepair attempts to repair any issues where a string literal is present in the query either unquoted or quoted incorrectly.  SQL requires strings to be surrounded with single quotes. Thus, SQLRepair removes double quotes and surrounds what appear to be unquoted string literals with single quotes.  %
The query is thus updated to \mintinline[breaklines]{sql}{SELECT item, price, quantity, country FROM fruitSellers WHERE country='US' AND quantity < 800}.

Resolving operator mismatch, column mismatch, and fixing strings resolves syntax errors, but often synthesis is needed to fully correct the semantic errors. 

\subsection{Synthesis Repair}
SQLRepair uses a SMT solver, Z3~\cite{z3} to synthesize parts of a query in need of repair~\cite{aprsurvey}. 
The synthesized parts, or patches, are composed of individual constants, operators, and column names.  The \texttt{(source, destination)} tables are used as test cases that must be simultaneously satisfied for a query to be successfully patched.

For each query, SQLRepair builds a system of constraints to represent the query logic. %
Given a set of example \texttt{(source, destination)} tables $E$ and a SQL query $q$, SQLRepair checks that:
$\forall e \in E, q \wedge source_e \rightarrow destination_e$. 
If the equation evaluates to true, Z3 returns $SAT$ and $q$ is correct; otherwise $q$ is incorrect and a candidate for repair. 

If $q$ is a repair candidate, SQLRepair inserts holes into $q$, for example by replacing a constant with  \mintinline[breaklines]{sql}{CONST_i}, forming $q'$, and provides $q'$ to the solver. 
If $q'$ is repairable by SQLRepair, Z3 returns $SAT$ and the solver has identified values for the holes in the satisfiable model. 
If $q'$ is not repairable by SQLRepair, Z3 returns $UNSAT$. 
SQLRepair supports five types of synthesis repairs.  After each repair stage, the process terminates if a successful repair can be made.  Repairs are performed in the following order:

\subsubsection{Constant Synthesis}
\label{sec:constantsynthesis}
For constants that are  compared to columns, SQLRepair replaces each constant in the \mintinline[breaklines]{sql}{WHERE} clause with \mintinline[breaklines]{sql}{CONST_i}. If a query contains \mintinline[breaklines]{sql}{CONST_1 OP_1 CONST_2}, SQLRepair does not replace either of the constants.  Synthesis is supported for integers and strings, although synthesized strings must be exact matches without wildcards.

\subsubsection{Operator Synthesis}
\label{sec:opsynthesis}
SQLRepair replaces each operator in $q$'s \mintinline[breaklines]{sql}{WHERE} clause with \mintinline[breaklines]{sql}{OP_j}.  SQLRepair supports synthesising operators for both string and integer types.  SQLRepair supports \mintinline[breaklines]{sql}{=} and \mintinline[breaklines]{sql}{!=} when dealing with strings, and \mintinline[breaklines]{sql}{=}, \mintinline[breaklines]{sql}{!=}, \mintinline[breaklines]{sql}{>}, \mintinline[breaklines]{sql}{>=}, \mintinline[breaklines]{sql}{<}, and \mintinline[breaklines]{sql}{<=} when dealing with integers.   %

\subsubsection{Column Synthesis}
\label{sec:colsynthesis}
SQLRepair inserts holes for the columns.  For example, a query $q = $
\ldots\mintinline[breaklines]{sql}{quantity OP_1 CONST_1} 
is replaced with $q' = $
\ldots\mintinline[breaklines]{sql}{COL_1 OP_1 CONST_1}, where \mintinline[breaklines]{sql}{COL_1} represents one of the columns in the source table. If SQLRepair fails to find a solution, column synthesis is repeated for each subclause in the original query, in order. %

\subsubsection{Clause Removal}
\label{sec:clauseremoval}
SQLRepair will remove subclauses one at a time to attempt a solution. For a query with \texttt{n} subclauses, if a correct solution cannot be found for \texttt{n} subclauses, but {can} be found with \texttt{1\ldots n-1} subclauses, SQLRepair will remove subsequent clauses that impede correctness.  If this step fails, the removed clauses are added back to the query before proceeding with \textit{Clause Synthesis}.

\subsubsection{Clause Synthesis}
\label{sec:clausesynthesis}
Some queries require additional \mintinline{sql}{WHERE} clauses or conditions. 
In this case, SQLRepair functions most similarly to Scythe~\cite{scythe}, and will synthesize new subclauses.  
Suppose in the column synthesis step, SQLRepair inserts holes such that $q' = $
\ldots\mintinline[breaklines]{sql}{WHERE COL_1 OP_1 CONST_1}, but is not able to find any columns, operators, and constant values that result in a solution.  At this point, SQLRepair attempts to make a repair by synthesizing in a new subclause.  More formally, SQLRepair will take a clause \ldots\mintinline[breaklines]{sql}{WHERE COL_1 OP_1 CONST_1} from the previous step, and add a new subclause, giving $q' = $
\ldots\mintinline[breaklines]{sql}{WHERE COL_1 OP_1 CONST_1 BOP_1 COL_2 OP_2 CONST_2}, where \mintinline[breaklines]{sql}{BOP_1} is a binary operator (\mintinline[breaklines]{sql}{AND} or \mintinline[breaklines]{sql}{OR}) and \mintinline[breaklines]{sql}{COL_2 OP_2 CONST_2} represents the abstracted form of a new subclause to be synthesized.   If values can be found, they are inserted into the query, and the repair is complete.  If no such values can be found, the query will be expanded again.  This process repeats until either a solution is found, or the query reaches the maximum of five subclauses, at which point the process is aborted and the repair is marked as failed.\footnote{In our experiment, the maximum number of added clauses in a successfully patched query was three.}

In the  example, after repairing \emph{Operator Mismatch} and \emph{Column Mismatch} and performing \emph{String Repair} the query: $q = $
\mintinline[breaklines]{sql}{SELECT item, price, quantity, country FROM fruitSellers WHERE country='US' AND quantity < 800} is incorrect. Thus, SQLRepair creates: $q' = $ 
\mintinline[breaklines]{sql}{SELECT item, price, quantity, country FROM fruitSellers WHERE country = 'US' AND quantity OP_1 CONST_1}. When Z3 returns $SAT$, SQLRepair uses the satisfiable model to replace \mintinline[breaklines]{sql}{OP_1} $ \rightarrow \mathrel{\mathtt{!=}}$ and \mintinline[breaklines]{sql}{CONST_1} $ \rightarrow 500$, creating a correct query.

\begin{table*}[]

\caption{
    Classifications of syntax errors introduced by students across both phases.  %
}
\vspace{-8pt}
\label{tab:synerror}
\begin{tabular*}{\textwidth}{l @{\extracolsep{\fill}} rrrll}
Error Type                         & \makecell{CS2 \\ Number (\%)} & \makecell{SE \\ Number (\%)} & \makecell{Total \\ Number (\%)} &  Example                                                \\
\toprule
Broken operator                      & 188  (27.6\%)   &  29 (21.2\%)   &  217 (26.5\%)    & \mintinline[breaklines]{sql}{SELECT RUI FROM bravo WHERE CUI1 == 'C0000039'}                        \\

Column reference error               & 118 (17.3\%)    &  16 (11.7\%)   &  134 (16.4\%)  & \makecell[l]{\mintinline[breaklines]{sql}{SELECT DISTINCT CUI FROM juliett, india WHERE}\\ \mintinline[breaklines]{sql}{juliett.CUI = india.CUI}} \\

Quotes on strings                    & 87 (12.8\%)     &  40 (29.2\%)   &  127 (15.5\%)   & \mintinline[breaklines]{sql}{SELECT * FROM foxtrot WHERE TTY = PT}                                  \\

Incomplete query                     & 91 (13.4\%)     &  6 (4.4\%)     &  97  (11.9\%) &  \mintinline[breaklines]{sql}{SELECT DISTINCT WHERE MRRANK_RANK < 384;}   \\

Wrong order                          & 83 (12.2\%)     &  14 (10.2\%)   &  97  (11.9\%) & \mintinline[breaklines]{sql}{select LAT, STT, ISPREF distinct from juliett}                         \\

Table reference error                & 68 (10.0\%)     &  16 (11.7\%)   &  84  (10.3\%)  & \makecell[l]{\mintinline[breaklines]{sql}{SELECT STT, ISPREF FROM juliett WHERE} \\ \mintinline[breaklines]{sql}{india.CUI = juliett.CUI}}                         \\

Extra commas                         & 62 (9.1\%)      &  13 (9.5\%)    &  75 (9.2\%)   & \mintinline[breaklines]{sql}{SELECT CUI1, RUI, FROM bravo WHERE CUI2 = 'C0364349'}                  \\

Missing commas                       & 20 (2.9\%)      &  9 (6.6\%)     &  29 (3.5\%)  & \mintinline[breaklines]{sql}{SELECT RSAB TFR CFR FROM delta WHERE TFR > 470}                  \\

Miscellaneous                        & 38 (5.6\%)      &  1 (0.7\%)     &  39 (4.8\%) & \mintinline[breaklines]{sql}{SELECT CUI, STN, TUI from hotelORDER BY TUI DESC}        \\ \bottomrule
\end{tabular*}

\vspace{12pt}

\caption{
    Classifications of semantic errors made by students across both phases.  %
}
\vspace{-8pt}
\begin{tabular*}{\textwidth}{l @{\extracolsep{\fill}} rrrll}
Error Type                         & \makecell{CS2 \\ Number (\%)} & \makecell{SE \\ Number (\%)} & \makecell{Total \\ Number (\%)} &  Example                                                \\
\toprule
Wrong subclauses in WHERE                    & 828 (69.9\%) &  406 (76.7\%)   & 1,234 (72.0\%) & \mintinline[breaklines]{sql}{SELECT * FROM charlie}                                             \\

\makecell[l]{Missing or extra operator\\ (GROUP BY, DISTINCT, etc)} & 369 (31.1\%)  &  122 (23.1\%)  & 491 (28.6\%)  & \makecell[l]{\mintinline[breaklines]{sql}{SELECT LAT, STT, ISPREF FROM juliett, india} \\ \mintinline[breaklines]{sql}{WHERE juliett.CUI = india.CUI GROUP BY LAT}} \\

Wrong values in WHERE                        & 241  (20.3\%) &  74 (14.0\%)   & 315 (18.4\%)  & \makecell[l]{\mintinline[breaklines]{sql}{SELECT DISTINCT SVER FROM golf WHERE} \\ \mintinline[breaklines]{sql}{SVER < 2000}}                               \\

Wrong ordering                                      & 209 (17.6\%)   &  68 (12.9\%)  & 277 (16.2\%)  & \makecell[l]{\mintinline[breaklines]{sql}{SELECT DISTINCT * FROM echo ORDER BY} \\ \mintinline[breaklines]{sql}{MRRANK_RANK DESC}}  \\

Column mismatch                                     & 70 (5.9\%)  &  46 (8.7\%)      & 116 (6.8\%)   & \makecell[l]{\mintinline[breaklines]{sql}{SELECT * FROM juliett a, india b WHERE} \\ \mintinline[breaklines]{sql}{a.CUI = b.CUI}}           \\

Wrong operator in WHERE                      & 83 (7\%)   &  27 (5.1\%)       & 110 (6.4\%)   & \makecell[l]{\mintinline[breaklines]{sql}{select LAT, STT, ISPREF from india a,} \\ \mintinline[breaklines]{sql}{juliett b where a.CUI = b.CUI AND a.CVF = 256}}    \\

Missing join (implicit or explicit)                 & 43 (3.6\%)   &  21 (4.0\%)     & 64 (3.7\%)    & \makecell[l]{\mintinline[breaklines]{sql}{SELECT LAT, STT, ISPREF from juliett} \\ \mintinline[breaklines]{sql}{where TS='S';}}                                     \\

Miscellaneous                                                & 31 (2.6\%)   &  3 (0.6\%)      &  34 (2.0\%)   & \mintinline[breaklines]{sql}{SELECT DISTINCT SVER FROM golf;}         \\ \bottomrule
\end{tabular*}
\label{tab:semerror}
\end{table*}

\subsection{Analysis}

To identify queries to repair for RQ2, we considered any query that had a syntax error or semantic error.  We report on what SQLRepair can fix from Phase~1 and Phase~2.  Unlike with error classification, as discussed in Section~\ref{sec:sqleval}, a repaired query could be counted towards both the synthesis and non-synthesis categories, depending on precisely what repair operations were performed.

With RQ3, we seek to understand the quality of the repairs produced by SQLRepair.  Students in Phase~2 were shown multiple queries simultaneously (see Figure~\ref{fig:votingscreen}) and asked to rate the understandability of each one on a seven-point Likert scale.  Because students were shown multiple queries simultaneously,  we are interested in the relative ratings given to each one.  Thus, we perform a series of paired Mann–Whitney U analyses to understand how queries from one category compare to queries from another category.  %

\section{Results}
\label{sec:results}

In this section, we present quantitative and qualitative results showing the types of errors students introduce (RQ1), the types of repairs by SQLRepair (RQ2), and the repair quality (RQ3).

\subsection{RQ1: SQL Mistakes}
\label{sec:sqlmistakes}

The students in SE were more successful at solving the problems than the students in CS2 (see Table~\ref{tab:breakdown}).  Among queries submitted by SE students, 12.4\% (94 of 760) were correct, compared to 7.8\% (157 of 2,022) from CS2 students.  Additionally, perhaps due to exposure to more programming languages, the SE students introduced syntax errors at a lower rate (20.6\% of all queries with errors, vs. 36.4\% among CS2 students).   We performed a test of two proportions and found that the difference in overall success rates between groups was statistically significant ($p<.001$).  
For this reason, results for students from each course are presented separately.

Table~\ref{tab:synerror} and Table~\ref{tab:semerror} show the syntax and semantic errors students introduced, respectively. Because individual queries can contain multiple errors, a query can be counted in more than one category.  Each row in the table shows one of the categories, how many queries had errors of that type, a corresponding percentage, and a representative example from the category.  For example, the first row of Table~\ref{tab:synerror} is our syntax error category of a \textit{Broken operator}; we saw 217 of these, representing 26.5\% of the 817 queries with syntax errors.  The query \mintinline[breaklines]{sql}{SELECT RUI FROM bravo WHERE CUI1 == 'C0000039'} was placed into the \textit{Broken operator} group because the query uses a \mintinline[breaklines]{sql}{==} where it should have used a \mintinline[breaklines]{sql}{=}.  %

We notice similarities between our categories in Table~\ref{tab:synerror} and those reported by Taipalus and Per\"{a}l\"{a}~\cite{sqlteaching}; for instance, we both observed \textit{a column reference error}, \textit{wrong ordering} of SQL keywords, and \textit{miscellaneous syntax errors}.  Likewise, there is overlap between our categories and those of Ahadi, et al.~\cite{syntaxmistakes}; the \textit{column reference error} rank high in both lists, and their general \textit{syntax error} category appears similar to our \textit{broken operator} category.  Unfortunately, because they do not offer examples of their categories it is impossible to map our categories to theirs precisely.
The types of semantic errors that we saw are shown in Table ~\ref{tab:semerror}.   The most common issue was \textit{Wrong subclauses in the \mintinline[breaklines]{sql}{WHERE} clause}; this is the first row in the table and was observed in 1,234, or 72\%, of queries.  The prevalence here indicates that students had difficulty precisely describing the rows they wanted to include.  A \textit{Missing or extra operator} was the second most common issue, particularly among students in CS2.   In contrast to our categories of semantic errors, which represent cases where the analyzed query returns an incorrect result, Brass and Goldberg~\cite{semantic} focus on queries that are correct but complicated or difficult to read.   There is, however, overlap between our categories and those of Taipalus and Per\"{a}l\"{a}~\cite{sqlteaching}, such as a \textit{missing join}.  In addition, their category of \textit{duplicate rows} is similar to ours of a \textit{Missing [or extra] operator}.

The breakdown of successful queries and submitted queries on each problem is shown in Table~\ref{tab:breakdownperproblem}.
We note that certain problem types proved to be particularly challenging.  For example, Problem 9, which necessitated use of a join, was widely attempted (with a total of 350 attempts from 42 different participants) but was solved correctly by only a single student.  Problems involving compound \mintinline{sql}{WHERE} clauses (Problems 5 \& 6) proved difficult as well, with less than a third of students managing to solve each one correctly.  The lower success rates on these problems compared to single-condition selects (Problems 1 \& 2) likewise suggests that students struggle with understanding the interactions between multiple columns.

\mybox{\textit{\textbf{RQ1 Summary:} Students made eight main types of syntax mistakes, including misusing an operator and ambiguity with referenced columns, and seven main types of semantic mistakes, including using the wrong column(s) in a \mintinline[breaklines]{sql}{WHERE} clause, using wrong  constants, and missing operators such as \mintinline[breaklines]{sql}{GROUP BY} or \mintinline[breaklines]{sql}{DISTINCT}. Joins and compound clauses proved difficult for all students.}}

\begin{table*}[ht]
\centering
\caption{
    Successes per problem and per course.  Each cell represents the ratio between the number of correct attempts and the total number of attempts.  Success per participant represents the ratio between the number of students who attempted the problem and the number of students who solved it successfully.  Success per attempt represents the sum of correct attempts to total attempts across CS2 and SE. 
}
\begin{tabular}{clrrrrr}
 & & \multicolumn{2}{c}{\textbf{Course}}  &                                                                           \\
   \makecell{\textbf{}\\  \textbf{Problem}} & \textbf{Major Concept} & CS2 & SE &  \makecell{\textbf{Success}\\  \textbf{per participant}} &  \makecell{\textbf{Success}\\  \textbf{per attempt}}   \\
\midrule
1 & Single-condition select       & 31/115 (27.0\%)          & 14/25 (56.0\%)       & 45/51 (88.2\%)         & 45/140 (32.1\%)          \\
2 & Select with projection        & 17/233 (7.3\%)           & 9/163 (5.5\%)        & 26/49 (53.1\%)         & 26/396 (6.6\%) \\
3 & Inequality                    & 17/140 (12.1\%)          & 13/55 (23.6\%)       & 30/41 (73.2\%)         & 30/195 (15.4\%)          \\
4 & Projection and inequality     & 23/145 (15.9\%)          & 12/39 (30.8\%)       & 35/41 (85.4\%)         & 35/184 (19.2\%) \\
5 & Compound select               & 4/304  (1.3\%)           & 7/101 (6.9\%)        & 11/51 (21.6\%)         & 11/405 (2.7\%) \\
6 & Compound select with AND      & 6/256  (2.3\%)           & 8/113 (7.1\%)        & 14/44 (31.8\%)         & 14/369 (3.8\%)      \\
7 & Distinct                      & 23/153 (15.0\%)          & 11/38 (28.9\%)       & 34/52 (65.4\%)         & 34/191 (17.8\%) \\
8 & Ordering                      & 19/201 (9.5\%)           & 12/80 (15.0\%)       & 31/49 (63.3\%)         & 31/281 (11.0\%) \\
9 & Joins                         & 1/280  (0.4\%)           & 0/70  (0.0\%)        & 1/42  (2.4\%)          & 1/350 (0.3\%) \\
10 & Grouping                     & 16/195 (8.2\%)           & 8/76  (10.5\%)       & 24/45 (53.3\%)         & 24/271 (8.9\%) \\      
\bottomrule
\makecell{\textbf{Successful} \\ \textbf{attempts}} & - & \makecell{157/2022 \\(7.8\%)} & \makecell{94/760 \\(12.4\%)}  &

\end{tabular}
\label{tab:breakdownperproblem}
\end{table*}

\subsection{RQ2: SQLRepair}

\label{sec:sqlrepair}

\begin{table*}[]
\caption{
    Types of complete repairs from SQLRepair.  Non-synthesis repairs are presented first, followed by synthesis repairs.  Each section is sorted by the total number of repairs; percentages are computed over the total number of repaired queries.  Because many successfully repaired queries contain two or more repairs, the totals in each column sum to more than 100\%. The identifier associated with each repair type corresponds to the description in Section~\ref{sec:implementation}.
}
\vspace{-8pt}
\begin{tabular}{c lrrrp{3.05in}}
& \textbf{Repair Type}   &             \makecell{\textbf{CS2} \\ \textbf{Number} (\%)} & \makecell{\textbf{SE} \\ \textbf{Number} (\%)} & \makecell{\textbf{Total} \\ \textbf{Number} (\%)} & \textbf{Representative Example}\\
\toprule

 \parbox[t]{0mm}{\multirow{5}{*}{\rotatebox[origin=c]{90}{Non-Synthesis}}} & Column Mismatch (\ref{sec:colmismatch})           & 67 (13.7\%)    & 40 (16.1\%)    & 107 (14.5\%) &  \mintinline[breaklines]{sql}{ SELECT CUI, TUI FROM}\dots $\rightarrow$ \mintinline[breaklines]{sql}{ SELECT CUI, TUI, STN FROM} \dots       \\

& String Repair (\ref{sec:strrepair})             & 33 (6.8\%)     & 42 (16.9\%)    & 75 (10.2\%) &  \dots\mintinline[breaklines]{sql}{WHERE CUI2 = C0364349} $\rightarrow$ \dots\mintinline[breaklines]{sql}{WHERE CUI2 = 'C0364349'} \\

& Operator Mismatch (\ref{sec:opmismatch})          & 28 (5.7\%)     & 0 (0\%)         & 28 (3.8\%) &  \dots\mintinline[breaklines]{sql}{WHERE min==0} $\rightarrow$ \dots\mintinline[breaklines]{sql}{WHERE min = 0} \\

\midrule

\parbox[t]{0mm}{\multirow{8}{*}{\rotatebox[origin=c]{90}{Synthesis}}} &Column Synthesis (\ref{sec:colsynthesis})          & 252 (51.6\%)   & 141 (56.6\%)   & 393 (53.3\%)  & \dots\mintinline[breaklines]{sql}{WHERE REL = 'RO';} $\rightarrow$ \dots\mintinline[breaklines]{sql}{WHERE CUI2 = 'C0364349';} \\

&Clause Removal (\ref{sec:clauseremoval})           & 109 (22.3\%)    & 98 (39.4\%)   & 207 (28.1\%) & \dots\mintinline[breaklines]{sql}{WHERE CUI2 = 'C0364349' OR REL = 'RO'} $\rightarrow$ \dots\mintinline[breaklines]{sql}{WHERE CUI2 = 'C0364349'} \\

&Clause Synthesis (\ref{sec:clausesynthesis})      & 131 (26.8\%)    & 65 (26.1\%)   & 196  (26.6\%) & \mintinline[breaklines]{sql}{SELECT CUI1, RUI FROM bravo} $\rightarrow$ \mintinline[breaklines]{sql}{SELECT CUI1, RUI FROM bravo WHERE CUI2 = 'C0364349';} \\

&Constant Synthesis (\ref{sec:constantsynthesis})        & 114 (23.4\%)    & 21 (8.4\%)    & 135 (18.3\%) &  \dots\mintinline[breaklines]{sql}{WHERE CFR < 1834} $\rightarrow$ \dots\mintinline[breaklines]{sql}{WHERE CFR < 1865}                                                        \\
&Operator Synthesis (\ref{sec:opsynthesis})        & 39 (8.0\%)      & 12 (4.8\%)    & 51 (6.9\%) &  \dots\mintinline[breaklines]{sql}{WHERE TFR < 1850}\dots $\rightarrow$ \dots\mintinline[breaklines]{sql}{WHERE TFR <= 1965}\dots                                            \\

\bottomrule

\end{tabular}
\label{tab:complete}
\end{table*}

Our evaluation dataset consists of 2,531 incorrect SQL queries.  Of these, SQLRepair was able to find a repair for 737, giving an overall repair success rate of 29.1\%.  The different types of repairs made are shown in Table~\ref{tab:complete}.  The table is organized based on the repair types: the first three correspond to the three non-synthesis repairs supported, and the last five to the synthesis repairs.  For example, the first row, \textit{Column Mismatch}, is described in Section~\ref{sec:colmismatch}; this repair is made to 67 (13.7\% of 488) queries from CS2 and 40 (16.1\% of 249) from SE, totaling 107 (14.5\% of 737) of all repaired queries. The representative example modifies the \mintinline[breaklines]{sql}{SELECT} clause to return three columns instead of two. 

\subsubsection{Repaired Queries}
The most common repair type observed was \textit{Column Synthesis} (made to 393, or 53.3\% of 737, queries), where SQLRepair synthesizes an expression using a new column, replacing an existing expression.  The second most common repair type observed was \textit{Clause Removal}, where SQLRepair identifies and removes a \mintinline[breaklines]{sql}{WHERE} subclause that results in incorrect output.  The third most common synthesis repair type is \textit{Clause Synthesis}, where a new subclause is generated for the \mintinline[breaklines]{sql}{WHERE} clause.  Together, these three repairs correspond to the very common \textit{Wrong subclauses in WHERE clause}  error observed across both classes (see Table~\ref{tab:semerror}), where the resolution is to add, fix, or remove an incorrect clause.

We also observe that while the majority of repairs performed (982 of 1,192 repairs, or 82.4\%, from Table~\ref{tab:complete}) involve a synthesis repair, non-synthesis repairs play an important part in success as well.  In 107 cases, our tool fixes a \textit{Column Mismatch error} by identifying a query that is returning the wrong set of columns and rewrites the \mintinline[breaklines]{sql}{SELECT} clause accordingly.
Although this is a non-synthesis repair, it fixes queries from the \textit{Column reference error} category in Table~\ref{tab:synerror} and the \textit{Column mismatch} category in Table~\ref{tab:semerror}, thus covering both syntax and semantic errors.  Fixing unquoted or misquoted string literals (\textit{String Repair}) and incorrect C/Java style operators (\textit{Operator Mismatch}) happen less often, but a fix from one of these categories is still made to 75 and 28 queries, respectively.  Additionally, making non-synthesis repairs also opens up new possibilities for synthesis repairs: queries must be well-formed for synthesis repair to proceed, and non-synthesis repair fixes some cases where they are not.

More often than not, repaired queries requires a combination of repair operations. In fact, 433 (58.8\%) of the successfully repaired queries contained multiple repair operations.  For example, the query \mintinline[breaklines]{sql}{select * from delta WHERE CFR < 1696} was repaired by fixing both the columns to return (a \textit{Column Mismatch} repair and changing the 1696 to an 1865 (a \textit{Constant Synthesis} repair).  %

\subsubsection{Not Repaired Queries}
The remaining 1794 queries that could not be repaired fall into two major categories:

\paragraph{Unsupported functionality}  Some functionality necessary to solve the problems in Table~\ref{tab:problems} is not supported in SQLRepair, such as \mintinline[breaklines]{sql}{GROUP BY} or joins. Students also used functionality that was neither necessary nor supported (such as \mintinline[breaklines]{sql}{BETWEEN} and \mintinline[breaklines]{sql}{LIMIT}), which  rendered their queries unfixable.  

\paragraph{Miscellaneous syntax errors}  SQLRepair can fix some but not all syntax errors.  Errors such as a misspelled SQL keyword (e.g., \dots\mintinline[breaklines]{sql}{GROUPED BY}\dots) clauses placed in the wrong order (e.g., \mintinline[breaklines]{sql}{SELECT * DISTINCT}\dots) are not fixed automatically by our tool.

\subsubsection{Performance} We tested the performance of SQLRepair on an Intel i7-6700HQ running Linux Mint 18.  Successful repairs are found in a median of 231 milliseconds (max: 1,602) and unsuccessful repairs in a median of 196 milliseconds (max: 1,912).

\mybox{\textit{\textbf{RQ2:} SQLRepair automatically fixes 29.1\% of student queries with errors, covering both syntax and semantic errors.}%
        }

\subsection{RQ3: Repair Quality}

To understand the quality of the repairs produced by SQLRepair, once students in Phase~2 found a solution for a problem (or gave up), we presented them with several alternative solutions (see Section~\ref{sec:phase2}).  Students rated each query on a scale of 1 (very difficult to understand) to 7 (very easy to understand) and optionally provided a free response rationale.  We received a total of 281 voting responses (CS2: 183, SE: 98) and 81 rationales (CS2: 50, SE: 31).

Each query was from one of four categories (Section~\ref{sec:phase2}): MyCorrectQuery (MCQ), MyRepairedQuery (MRQ), OtherCorrectQuery (OCQ), OtherRepairedQuery (ORQ). On average, students found their own queries (MCQ) to be more understandable than their repaired queries (MRQ) (5.58 vs. 5.35, see Table~\ref{tab:votes}), but the difference is not significant. Thus, a student's repaired query could be used as an alternate way to solve a problem without sacrificing understandability, a divergence from prior work in automated program repair suggesting that machine-repaired code is less understandable than human-written code~\cite{understandability}.  %

Looking more closely at the data, a pairwise analysis can determine the within-participant differences in understandability between each query category. Using the four query categories, we  ran six paired Mann–Whitney U tests.  As all the p-values are above $0.1$, the data show that there is no statistical difference in understandability between human-written and machine-repaired SQL queries. For example, comparing the student's own query (MCQ) with their repaired queries (MRQ) yielded $p=0.662$. While there is a 0.23 difference in averages, representing a quarter of a level on the 7-point Likert scale, the difference is not significant. Comparing a student's own correct query (MCQ) against a correct query written by others (OCQ) also reveals no difference in understandability. Therefore, we find evidence that repaired queries and queries written by others are all viable candidates for presenting students with alternate implementations of SQL queries. 

Qualitatively, we observe that some students preferred their own solutions over all others; we received written responses such as ``\textit{I literally wrote [this query]}'' and ``\textit{I chose [this query] because it was exactly my solution}''.  However, this was not universally the case: one student remarked ``\textit{I can't believe how [bad] my answer is}''.  This suggests that automated repair can help students identify better solutions even after solving a problem correctly.

\begin{table}[tb]
\centering
\caption{
    Average Likert-scale understandability scores per course and per query type, where 1 maps to Very Difficult and 7 maps to Very Easy; 4 is a neutral response (neither easy nor difficult). Query categories are defined in Section~\ref{sec:phase2}.
}
\begin{tabular}{lrrr}
 & \multicolumn{2}{c}{\textbf{Course}}  &                \\
 & CS2             & SE & \makecell{\textbf{Overall}}   \\
\midrule
MyCorrectQuery (MCQ)        & 5.62                   & 5.54  &  5.58           \\
MyRepairedQuery (MRQ)      & 5.32                   & 5.41  &  5.35           \\
OtherCorrectQuery (OCQ)    & 5.04                   & 5.41  &  5.17           \\
OtherRepairedQuery (ORQ)   & 5.03                   & 5.38  &  5.15           \\
\end{tabular}
\label{tab:votes}
\end{table}

\mybox{\textit{\textbf{RQ3:} Queries repaired by SQLRepair are rated as equal in understandability compared to queries written by the students themselves, suggesting repaired queries could be useful for presenting students with alternate queries.}}

\section{Discussion}
\label{sec:discussion}

Here, we discuss the implications of our results, present opportunities for future work, and discuss threats to validity.

\subsection{Implications}
\label{sec:imp}

By analyzing the SQL queries written by students new to SQL, we see that certain topics are particularly challenging; most students struggled with joins, ordering, and compound clauses.  When SQL is used, it is typically with more than one table, so teaching joins is a necessity~\cite{sqlusage}.  By contrast, students had less difficulty with operators such as \mintinline{sql}{GROUP BY}.  All concepts were introduced to students in a similar way, with background information provided through slides and live examples showing how they work in practice.  These results suggest that some topics remained more difficult for students to understand and thus may require additional instruction.  %

To the best of our knowledge, SQLRepair is the first automated repair (APR) tool for SQL queries, and our results provide preliminary evidence that this can be useful in education.  Patitsas, et al. report that presenting students with multiple solutions side-by-side can improve learning outcomes~\cite{compcont}.  In cases where peer instruction is unavailable, our results suggest repair tools may be able to provide alternative solutions for students to visualize.  Providing hints or iterative refinement rather than just a new solution may further improve the process.

While the overall repair rate of SQLRepair is lower than many general-purpose repair tools, this is a first step and the availability of our dataset should allow future SQL repair tools to improve on our efforts reported here for educational and professional audiences. %

\subsection{Future Work}
We have identified several promising directions for future work in program repair to support learners.%

The single most challenging problem for students  was one that involved joining two different tables together on a common column.  This suggests that students struggle to see the big picture and how their data connects together.  Tools such as MySQL Workbench allow reverse-engineering an entity-relation diagram from an existing database schema, and the produced diagrams can be used much like UML class diagrams to introduce new developers to an existing design.  A database IDE that automatically shows the relationship between tables when two or more are included in a query could help users see and utilise the connections in their data.

We observed, and several students affirmed in their comments, that it is challenging to identify patterns within a table and thus pick out desired rows (i.e., forming queries from examples is challenging).  Tooling that highlights similarities and differences between selected columns of two or more rows could help the user better identify relevant patterns.

More generally, our results suggest that program repair may be a useful educational tool for presenting alternate solutions to a problem. 
Notebooks such as Jupyter have become a popular way for performing exploratory data analysis, particularly among end-user programmers, because they allow intermingling code, written descriptions, and results~\cite{notebooks}.  While most such notebooks focus on Python or R, SQL has a place within the data science world as well, and integrating synthesis or repair tools could help make the learning process easier for many students.

All of the repairs produced by SQLRepair follow the steps listed in Section~\ref{sec:implementation}.  The order in which repairs are performed has the potential to impact the query that is ultimately produced.  Future work could study whether performing repairs in a different order impacts the quality of the query produced by potentially producing more concise or understandable solutions.

\subsection{Threats to Validity}
\label{sec:threats}

Our conclusions may not generalize to different student body populations.  The students who signed up to participate for our Phase~2 evaluations did so on the promise of extra credit.  Consequently, there may be a selection bias.  %

The problems we had students complete were based off of the UMLS dataset; %
it is unknown whether the nature of the dataset contributed to the difficulty students faced when solving problems.  The specific errors that students faced may not generalize to different problems.  It is possible that the context of the data made problems more difficult than if students had been working with more familiar data.  However, we expect the data to be equally unfamiliar to all students.

In this work, we use the understandability rating that a student gives a query as a proxy for the quality of the query that has been produced.  However, this merely asks students to read the query and then offer a vote on it; we do not ask them to integrate the queries produced into a larger application or modify the query to solve a problem that is similar but not identical.  Consequently, students who are using the queries in a different context may have different priorities for what makes a query understandable or not.

\section{Related Work}
\label{sec:relwork}

Existing work in teaching SQL to undergraduates focuses on how students learn SQL~\cite{sqllearning, sqlhomework} and the types of semantic and syntax errors made~\cite{syntaxmistakes,semantic,sqlteaching}.  Migler and Dekhtyar~\cite{sqllearning} break down the exercises students solve in an undergraduate databases course around their primary concept, and find that joins and subqueries are the most challenging. Our observations agree with theirs in that students have a harder time solving problems involving multiple tables.  

Poulsen, et al.~\cite{sqlhomework} study SQL queries students write in an upper-level databases course and find persistent issues with not just difficult semantic concepts such as nested queries and grouping, but syntax errors.  Ahadi, et al.~\cite{syntaxmistakes} consider only syntax errors; we observe that students make significantly more semantic errors than syntax errors. Brass and Goldberg~\cite{semantic} present a list of semantic errors in SQL queries, showing some errors guarantee an incorrect result and others produce a query that is substantially more complicated than is necessary.  However, their work does not assess error frequency, which we report in Section~\ref{sec:sqlmistakes}, and therefore cannot be used as a basis for direct comparison. %
Most similar to our work, Taipalus and Per\"{a}l\"{a}~\cite{sqlteaching} and Taipalus, et al.~\cite{sqlerrorscomplications} present a breakdown of errors made into semantic and syntax categories.  However, their student population is from a more advanced databases course. In our work, we offer a similar breakdown for  novice students.

Weise, et al.~\cite{styleComprehension, styleComprehension2} study student preferences for Java and Python code written in different programming styles, and their ability to understand code written in an ``expert'' style.  They find that many students prefer a more naive, or verbose, approach, but are capable of understanding code that uses more expert approaches.  Similarly, our work asks students to choose between several different queries, which may be more-or-less expertly written, to select the one that is easiest to understand.  Maalej, et al.~\cite{programComprehension} study how professional developers comprehend and understand the code they are working with.  While we do not ask them to explain their comprehension process, we nonetheless expect them to perform many of the same steps by reading and comparing multiple queries.

Stolee and Elbaum~\cite{stolee} demonstrate that students can take provided SQL queries and write corresponding input-output table pairs for them.  Our work asks them to do the reverse.

To the best of our knowledge, SQLRepair represents the first application of automated program repair to SQL.
However, some research efforts have produced tools for SQL query construction using program synthesis. 
SCYTHE~\cite{scythe} takes input-output examples and generates a query capable of performing the transformation,  %
but is limited in that it does not  support common operators such as projection.  Finally, SCYTHE supports only a single \texttt{(source, destination)} pair, while SQLRepair supports arbitrarily many.

Existing work by Solar-Lezama~\cite{synthesis} in program synthesis by sketching demonstrates that it is feasible to provide \textit{part} of a program, and to have automated tools fill in the remainder of it.  This is the approach that we use for synthesis repairs.

Drosos, et al.~\cite{drososwrexpysynthesis} present a tool, Wrex, for performing program synthesis in Jupyter notebooks.  They focus on producing Python code that is easy for humans to read and understand so the code is more likely to be used going forwards.  We report results similar to theirs, showing that code synthesized by a tool is of sufficiently high quality to be used.  %

\section{Conclusion}
\label{sec:conclusion}

In this work, we have analyzed the mistakes that undergraduate students make when working with SQL for the first time by studying the errors they introduce.  We found that the majority of queries contain one or more syntax or semantic error, and that semantic errors make up a majority of errors introduced.  We found that junior-level students perform better than sophomore-level students, solving more problems correctly and introducing syntax errors at a lower rate.  Among the more advanced SQL topics covered, students particularly struggle with joins, thus suggesting a need for teaching students to see and utilise patterns in data.  We have also demonstrated that SQLRepair can fix 29.1\% of queries with errors.  By demonstrating that APR techniques are applicable to SQL, we pave the way for additional automated repair of special-purpose programming languages.  Finally, our results suggest that automated repair may support students as they learn SQL.  Students rate our tool-produced repairs as good as queries written by themselves or other students, and thus automated repairs may make a compelling teaching tool when peer instruction and feedback is unavailable.

\section*{Acknowledgments}

This work was supported in part by NSF SHF grants \#1645136 and \#1749936.  We would like to thank Gina R. Bai for her comments on this work and the students of NC State University's Summer 2019 CSC 216 course and Fall 2020 CSC 216 and CSC 326 courses for allowing us to use their data for analysis and evaluation.

\section*{Data Availability}
All queries collected, SQLRepair, and supporting tools for analysis are available on Zenodo.

\bibliographystyle{IEEEtran}
\bibliography{bibfile}

\end{document}